\documentclass[aps, pra, showkeys, reprint, nofootinbib, longbibliography]{revtex4-1} 

\usepackage[usenames,dvipsnames]{color} 
\usepackage{amsmath} 
\usepackage{txfonts} 
\usepackage{upgreek} 
\usepackage{todonotes} 
\usepackage{amssymb} 
\usepackage{dsfont} 
\usepackage{braket} 
\usepackage{array} 
\usepackage[unicode,breaklinks]{hyperref}
\hypersetup{
    unicode=true,
    a4paper=true,
    plainpages=false,
    pdftitle={An epsilon-pseudoclassical Model for Quantum Resonances in a Periodically Laser-Driven Dilute Atomic Gas},
    pdfauthor={B. J. Beswick, S. A. Gardiner, I. G. Hughes, H. P. A. G. Astier, M. F. Andersen},
    pdfsubject={An epsilon-pseudoclassical Model for Quantum Resonances in a Periodically Laser-Driven Dilute Atomic Gas},
    colorlinks=true,
    linkcolor=blue,
    citecolor=blue,
    filecolor=black,
    urlcolor=blue
}
\hypersetup{colorlinks=true,pdfborder={0 0 0}} 
\usepackage{upgreek} 

\usepackage[caption=false]{subfig} 
\usepackage{graphicx} 
\usepackage[countmax]{subfloat}
\usepackage{import} 
\usepackage{transparent} 

\definecolor{MyGreen}{rgb}{0,0.5,0} 

\graphicspath{ {../Figures/} }

\begin{document}

\title{An $\boldsymbol{\epsilon}$-pseudoclassical model for quantum resonances in a cold dilute atomic gas periodically driven by finite-duration standing-wave laser pulses}
\author{Benjamin T. Beswick}\email{b.t.beswick@durham.ac.uk} 
\author{Ifan G. Hughes}
\author{Simon A. Gardiner} \email{s.a.gardiner@durham.ac.uk}
\affiliation{Joint Quantum Centre (JQC) Durham--Newcastle,
Department of Physics, Durham University, Durham DH1 3LE, United Kingdom}
\author{Hippolyte P. A. G. Astier} \email{hpaga2@cam.ac.uk} \affiliation{Department of Physics,
Cavendish Laboratory,
Cambridge
CB3 0HE, United Kingdom}
\author{Mikkel F. Andersen}
\affiliation{Dodd--Walls Centre for Photonics and Quantum Technologies, Department of Physics, University of Otago,  Dunedin 9016, New Zealand}
\author{Boris Daszuta}
\affiliation{Department of Mathematics and Statistics, University of Otago, 
Dunedin 9054, New Zealand}
\date{\today}

\begin{abstract}
Atom interferometers are a useful tool for precision measurements of fundamental physical phenomena, ranging from local gravitational field strength to the atomic fine structure constant. In such experiments, it is desirable to implement a high momentum transfer ``beam-splitter,'' which may be achieved by inducing quantum resonance in a finite-temperature laser-driven atomic gas. We use Monte Carlo simulations to investigate these quantum resonances in the regime where the gas receives laser pulses of finite duration, and demonstrate that an $\epsilon$-classical model for the dynamics of the gas atoms is capable of reproducing quantum resonant behavior for both zero-temperature and finite-temperature non-interacting gases. We show that this model agrees well with the fully quantum treatment of the system over a time-scale set by the choice of experimental parameters. We also show that this model is capable of correctly treating the time-reversal mechanism necessary for implementing an interferometer with this physical configuration.
\end{abstract}

\maketitle

\section{Introduction}
Microkelvin-temperature cold-atom-gases are a useful medium for atom-optical experiments, including atom interferometry \cite{atom_interferometry_2006}. For light-pulse atom-interferometry experiments it is desirable to implement a high momentum transfer ``beam splitter'' \cite{Clade_Pierre_2009, Muller_Holger_2009, Choiw_Sheng-wey_2011}, which can be realized by subjecting an atomic gas to a periodically-pulsed-optical-standing-wave. By tuning the period of the pulse sequence to a specific value known as the Talbot time, the phenomenon of \textit{quantum resonance} can be exploited to coherently split the atomic population of the gas in momentum space using minimal laser power. 

A dilute atomic gas receiving pulses of ``short'' duration is well approximated by the atom-optical $\delta$-kicked rotor Hamiltonian \cite{saunders_2009}. The atom-optical $\delta$-kicked-rotor has long been the subject of study in the field of quantum chaos \cite{the_transition_to_chaos_reichl,Regular_and_Chaotic_Dynamics}, aided by the relative simplicity of the both the classical and quantum $\delta$-kicked rotor. This includes the existence of some analytical results, as well as the ease with which the quantum $\delta$-kicked rotor lends itself to Fourier methods \cite{quantum_resonance_periodic_field,Doherty_2000_momentum_distributions_simulation}. Though laser pulses of truly infinitesimal duration are clearly unachievable experimentally, this model successfully describes experiments where the distance traveled by the atomic center of mass during each pulse is negligible relative to the spatial period of the standing wave \cite{Moore_1995_atom_optics_realization,kicked_rotor_wigner,phase_noise_dkr_white_ruddell_hoogerland,darcy_2001,darcy_2001_momentum_diffusion,Sadgrove_2004,Bharucha_1999_dynamical_localization,Moore_1994_dynamical_localization,Klappauf_1999_dynamical_localization-cesium,Steck_2000_dynamical_localization,Milner_2000_dynamical_localization,Oskay_2003_dynamical_localization,Vant_2000_dynamical_localization,Kanem_2007_higher_order_quantum_resonances,Duffy_2004_DKHO_BEC,
Behinaein_2006_accelerator_modes,Ryu_2006_higher_order_quantum_resonances_BEC,Szriftgiser_2002_sub_fourier_resonances,Ammann_1998_dynamical_localization_excited_state_mixing,Vant_1999_dkr_cantori_cesium,
Williams_2004_diffusion_resonances,Duffy_2004_early_time_diffusion_BEC} (the so called Raman--Nath regime \cite{daszuta_andersen_2012}). However, experiments indicate that finite pulse-duration effects can increase the sensitivity of atom interferometry experiments  \cite{lattice_interferometer_andersen}. This consideration, coupled with the fact that the infinitesimal pulse approach gives erroneous predictions over larger time-scales \cite{Oskay_2000}, motivates their incorporation into the kicked particle Hamiltonian. Though finite duration pulse atom interferometers have been investigated numerically for a single kicked particle \cite{daszuta_andersen_2012}, an investigation for a thermal gas of kicked particles is absent from the literature. 

A possible reason for this absence is that simulating driven systems with finite-duration pulses is notably more numerically complex than simulating systems with $\delta$-kicks \cite{daszuta_andersen_2012}, and this problem scales substantially with the number of particles. Given that knowledge of how the momentum distribution changes over time is necessary for designing and operating light-pulse atom-interferometry experiments, we are motivated to introduce a computationally simpler model, which can give accurate results for a typical experimental set-up.

In this paper we introduce an $\epsilon$-pseudoclassical model for the quantum kicked particle conceptually similar to that introduced to describe quantum accelerator modes by Fishman, Guarneri and Rebbuzzini \cite{fishman_guarneri_2003, stable_resonances_fishman}. This model is attractive due to its mathematical simplicity and the minimal computational complexity of the numerics. We explore the predictions of this model using a Monte Carlo approach, and compare the results to a fully quantum treatment. We find that the model captures the essential features of quantum resonant dynamics in finite-temperature driven gases.
\begin{figure}[!ht]
{\centering
\includegraphics{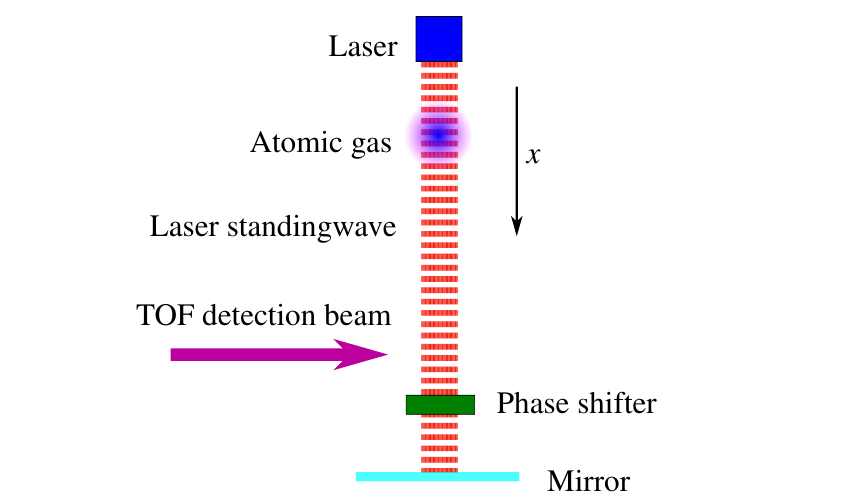}
\caption{\label{fig:drivengasdiagram} Schematic of a possible experimental setup \cite{godun_2000_accelerator_modes}.  The Time-Of-Flight (TOF) beam measures the atomic momentum distribution. If vertically oriented, the effect of the gravitational field can be transformed away \cite{saunders_halkyard_gardiner_challis_2009} by use of the phase shifter element, for example an electro-optic modulator \cite{godun_2000_accelerator_modes}.
}
}
\end{figure}

The paper is organised as follows:
in section \ref{overview} we overview experimental considerations, and describe the model system Hamiltonian and the time-evolution it generates;
in section \ref{finite_duration} we derive how to treat the existence of finite-duration pulse (assuming we are in the equivalent to a quantum-resonant regime for the $\delta$-kicked rotor) using an $\epsilon$-pseudoclassical model;
in section \ref{monte_carlo} we describe the Monte Carlo methodologies we use to determine our numerical results;
in section \ref{results} we compare and contrast numerical results using both full quantum dynamics and the pseudoclassical model;
and in section \ref{conclusions} we present our conclusions.

\section{System overview\label{overview}}
\subsection{Experimental considerations\label{experiment}}
As a typical system, one can consider a cloud of $10^5$ Cesium 133 atoms. This can be relatively straightforwardly confined and cooled in a MOT (magneto-optical trap), followed by an optical molasses, to a temperature of $\sim 5\mu \mathrm{K}$. In such a regime the resulting cold-atom gas is sufficiently dilute that atom--atom interactions can typically be neglected. Even lower temperatures can be achieved by Raman-sideband-cooling \cite{kasevich_chu_1992}, or by cooling to quantum degeneracy \cite{Duffy_2004_DKHO_BEC,Duffy_2004_early_time_diffusion_BEC}
(inter-atomic interactions can be significant within a Bose--Einstein condensate, however these can in principle be substantially tuned away by exploiting an appropriate magnetic Feshbach resonance \cite{inouye_andrews_1998,roberts_claussen_1998,kohler_goral_2006,gustavsson_haller_2008,molony_gregory_2014}, or letting the cloud expand). 

The atomic cloud can then be released under gravity, while two counter-propagating laser beams of wavelength $\lambda_L$ (choosing $\lambda_L=852\mathrm{nm}$ corresponds to the wavelength of the cesium $\mathrm{D}_2$ transition) form a laser standing wave in the horizontal direction (see Fig.~\ref{fig:drivengasdiagram}), which can be periodically pulsed \cite{kicked_rotor_wigner,phase_noise_dkr_white_ruddell_hoogerland,darcy_2001,darcy_2001_momentum_diffusion,Sadgrove_2004,Bharucha_1999_dynamical_localization,Moore_1994_dynamical_localization,Klappauf_1999_dynamical_localization-cesium,Steck_2000_dynamical_localization,Milner_2000_dynamical_localization,Oskay_2003_dynamical_localization,Vant_2000_dynamical_localization,Kanem_2007_higher_order_quantum_resonances,Duffy_2004_DKHO_BEC,
Behinaein_2006_accelerator_modes,Ryu_2006_higher_order_quantum_resonances_BEC,Szriftgiser_2002_sub_fourier_resonances,Ammann_1998_dynamical_localization_excited_state_mixing,Vant_1999_dkr_cantori_cesium,
Williams_2004_diffusion_resonances,Duffy_2004_early_time_diffusion_BEC}. By carefully tuning the phase-shifter element in Fig.~\ref{fig:drivengasdiagram}, the laser beams will form a ``walking wave,'' appearing as a standing wave in a frame comoving with the local gravitational acceleration \cite{saunders_halkyard_gardiner_challis_2009,godun_2000_accelerator_modes}. Neglecting interactions allows for a theoretical description using a single-particle Hamiltonian, which we describe in section \ref{Hamiltonian}.

After receiving a set number of laser pulses, a time-of-flight measurement can be performed to determine the momentum distribution of the gas (and thence its momentum variance). These experimental observables are typically what one would measure in light-pulse atom-interferometry experiments (see section \ref{quantum_resonance}), and we explain how they may be predicted numerically in section \ref{monte_carlo}.

\subsection{System Hamiltonian\label{Hamiltonian}}
During a laser pulse, the appropriate single-particle Hamiltonian describes a two-level atom (ground state $|g\rangle$ and excited state $|e \rangle$) of mass $M$  coupled to a laser standing wave of angular frequency $\omega_L$, wavenumber $k_L \equiv 2 \pi / \lambda_L$, and phase $\phi$ \cite{saunders_halkyard_challis_gardiner_2007,zheng}:
\begin{equation}
\begin{split}
\hat{H}_{\mathrm{2L}} =& \frac{\hbar\omega_0}{2} \left ( | e \rangle \langle e | - | g \rangle \langle g | \right)  + \frac{\hat{p}^2}{2M} 
\\& + 
\frac{\hbar \Omega}{2} \cos(k_L \hat{x}) \left[e^{-i(\omega_L t - \phi)} | e \rangle \langle g | +\mathrm{H.c.} \right]\, ,
\end{split}
\label{2levelatominlaserfield}
\end{equation}
where $\Omega$ is the on-resonance Rabi frequency,  $t$ is the time, and H.c.\ stands for Hermitian conjugate. Here, $\hat{x}$ and $\hat{p}$ represent the atomic position and momentum 
along the axis of the laser standing wave.\footnote{We may consider the center-of-mass dynamics in the $x$ direction in isolation, as they separate from the remaining center-of-mass degrees of freedom.} Transforming to an appropriate rotating frame, and adiabatically eliminating the excited state (assuming the laser field to be far-detuned and that all population begins in the ground state also justifies our neglect of spontaneous emission) results in the Hamiltonian \cite{saunders_halkyard_challis_gardiner_2007}
\begin{equation}
\hat{H}_{\mathrm{2L}}'' = \frac{\hat{p}^2}{2M} - \frac{\hbar \Omega^2}{8\Delta} \cos(2k_L \hat{x}) ,
\label{2levelatominlaserfield''}
\end{equation}
where we have defined\footnote{Note that the detuning is usually defined as equal to $\omega_L-\omega_0$ \cite{Foot} and thus is equal to $-\Delta$ as defined in this paper. Within the context of atom-optical $\delta$-kicked rotors the convention used in this paper is typical, however \cite{Moore_1994_dynamical_localization, Moore_1995_atom_optics_realization}.} $\Delta \equiv \omega_0 - \omega_L$. We describe the standing wave being periodically switched on and off through the dimensionless time-dependent function $f(t)$, giving
\begin{equation}
\hat{H} = \frac{\hat{p}^2}{2M} - \hbar \phi_{d} \cos(K\hat{x})
\frac{f(t)}{t_{p}},
\label{kickedparticleHfinite}
\end{equation}
where we have introduced $K \equiv 2k_L$ and $\phi_d \equiv \Omega^2 t_{p} /8\Delta$. The function
$f(t) = \sum_{n=-\infty}^{\infty}F_{\mathrm{sq}}(t-nT,t_{p})$, where 
\begin{equation}
F_{\mathrm{sq}}(t,t_{p})=
\begin{cases}
1& \text{for } 0 < t \leq t_{p}, \\
0& \text{for }  t\leq 0 \text{ or } t>t_{p}.
\end{cases}
\label{squarepulsetrain}
\end{equation}
describes a square pulse of duration $t_{p}$. This is typically a reasonable description of atom optical experiments \cite{Klappauf_1999_dynamical_localization-cesium}. As  $t_{p} \to 0$, then $f(t)/t_{p} \to \sum_{n=-\infty}^{\infty} \delta (t-nT)$, and in this limit Eq.~(\ref{kickedparticleHfinite}) reduces to the familiar $\delta$-kicked particle Hamiltonian described in \cite{saunders_halkyard_challis_gardiner_2007}.

\subsection{Time evolution \label{time_evolution}}
The time-periodicity of the Hamiltonian allows us to define a Floquet operator $\hat{F}$, such that $|\psi_{n+1}\rangle = \hat{F} |\psi_n\rangle$, where $|\psi_n\rangle$ denotes the state of the system immediately before the $n^\mathrm{th}$ kick: 
\begin{equation}
\begin{split}
\hat{F}=
\hat{U}_\mathrm{Free}\hat{U}_\mathrm{Kick}
 = &  \exp
\left(
-i
\frac{ \hat{p}^{2}}{2M}
\frac{[T-t_{p}]}{\hbar}
\right)
\\  & \times
\exp
\left(
-i\left[
\frac{ \hat{p}^{2}}{2M}
-\frac{\hbar \phi_d}{t_{p}}\cos(K\hat{x})\right]
\frac{t_{p}}{\hbar}
\right),
\end{split}
\label{floquetfiniteduration}
\end{equation}
where $\hat{U}_\mathrm{Free}$ governs the ``between-kick'' free evolution, and $\hat{U}_\mathrm{Kick}$ governs the time evolution while the kick is applied. 

It is convenient to partition the position and momentum operators \cite{bach_burnett_d'arcy_gardiner_2005}, such that:
\begin{subequations}
\begin{align}
K\hat{x} & = 2\pi\hat{l} + \hat{\theta}, \\
\hat{l}|Kx = 2\pi l + \theta\rangle & = l|Kx = 2\pi l + \theta\rangle, \\
\hat{\theta}|Kx = 2\pi l + \theta\rangle & =  \theta|Kx = 2\pi l + \theta\rangle,
\end{align}
\end{subequations}
where $l \in \mathbb{Z}$ and $\theta \in \left[0,2\pi \right)$ is effectively an angle variable; and 
\begin{subequations}
\begin{align}
(\hbar K)^{-1}\hat{p}  & = \hat{k} + \hat{\beta},
\label{Eq:MomentumParam}
\\
\hat{k}|(\hbar K)^{-1}p = k + \beta\rangle 
& = k|(\hbar K)^{-1}p = k + \beta\rangle,
\label{keigenstates}
\\
\hat{\beta}|(\hbar K)^{-1}p = k + \beta\rangle 
& = \beta|(\hbar K)^{-1}p = k + \beta\rangle,
\label{kbetaeigenstates}
\end{align}
\end{subequations}
with $k \in \mathbb{Z}$ and $\beta \in [-1/2,1/2)$. We can speak of $k$ as the discrete part of the dimensionless momentum $(\hbar K)^{-1}p$, and $\beta$ as the continuous part or \textit{quasimomentum}.

Fourier analysis of the Floquet operator $\hat{F}$ reveals that only momentum states separated by integer multiples of $\hbar K$ are coupled \cite{darcy_2001}, and so $\beta$ must be a conserved quantity; in other words $[\hat{\beta},\hat{H}]=0$ \cite{kicked_rotor_wigner,bach_burnett_d'arcy_gardiner_2005}. Within any specified quasimomentum subspace we can therefore consider the time evolution to be governed by
\begin{equation}
\begin{split}
\hat{F}(\beta) =& \exp
\left(
-i
\frac{ [\hbar K(\hat{k}+\beta)]^{2}}{2M}
\frac{[T-t_{p}]}{\hbar}
\right)
\\& \times
\exp
\left(
-i\left\{
\frac{ [\hbar K(\hat{k}+\beta)]^{2}}{2M}
-\frac{\hbar \phi_d}{t_{p}}\cos(\hat{\theta})\right\}\frac{t_{p}}{\hbar}
\right)\, .
\end{split}
\label{floquetfinitedurationB}
\end{equation}
We now have a continuum of Floquet operators, one for each $\beta$ subspace, within which $\beta$ can be considered simply a number \cite{fishman_guarneri_2003, stable_resonances_fishman,bach_burnett_d'arcy_gardiner_2005}. For the most general time evolutions one should in principle take relative phases between these subspaces into account, however this can be neglected if we do not consider coherent superpositions of states with different values of $\beta$.

\subsection{Quantum resonance, antiresonance and time-reversal\label{quantum_resonance}}
For the $\delta$-kicked rotor, quantum resonance occurs when the free evolution between kicks has no net effect on the state of the system \cite{saunders_2009,quantum_resonance_periodic_field,kicked_rotor_wigner,delocalization_shepelyansky,the_transition_to_chaos_reichl,power_law_behaviour_Halkyard}.  Referring to Eq.~(\ref{floquetfinitedurationB}) when $\beta=0$ and $t_{p}\rightarrow 0$, this corresponds formally to requiring $\hat{U}_{\mathrm{Free}}$ to collapse to the identity operator. Recalling that $\hat{k}$ has integer eigenvalues, this is fulfilled when
\begin{equation}
T= T_T\equiv \frac{4\pi M}{\hbar K^2}\, ,
\label{talbottime}
\end{equation}
or any integer multiple thereof. The quantity $T_T$ is known as the \textit{Talbot time\/} \cite{Oberthaler_1999_accelerator_modes,godun_2000_accelerator_modes}, in analogy with the \textit{Talbot length\/} of optics \cite{hecht_2002}. Within the $\beta = 0$ subspace (which maps exactly to the case of the quantum $\delta$-kicked rotor, with its  intrinsically discrete angular momentum spectrum), adjusting the period to an integer multiple of the Talbot time gives rise to an exactly quadratic increase in $\langle \hat{p}^2 \rangle$ over time, given by $\langle \hat{p}^2 \rangle_n=\hbar^2 K^2 \phi_d^2 n^2/2$ \cite{power_law_behaviour_Halkyard,BEC_Ullah}, where $n$ is the number of kicks.

Assuming the initial momentum distribution is symmetric about a mean value of zero, such ballistic growth of the system energy occurs via significant population being transferred into high-magnitude momentum states of opposite value (leading, at low temperatures, to a distribution with large, negative \textit{kurtosis\/} \cite{power_law_behaviour_Halkyard}). This splitting of the atomic momentum-distribution can form the first component of a light-pulse atom-interferometer \cite{Cronin_2009_atom_interferometry,daszuta_andersen_2012},
acting as the atom-optical analogue of a beam-splitter in classical optics. In an interferometric experiment, a relative phase would be accumulated between the ``arms'' of the resultant split cloud, due to coherent evolution caused by a perturbation to be measured. At a time $t_\mathrm{R}$, the laser standing-wave can be near-instantaneously phase-shifted in $\theta$ by an offset of $\pi$, which effectively reverses the quantum resonant dynamics, and causing the momentum-state populations to recombine some time later. At this time the relative phase can be extracted, and hence the magnitude of the perturbation.

For the case where the period $T$ is set to a half integer multiple of the Talbot time a phenomenon known as antiresonance can also be observed, characterized by kick-to-kick motion where there is no net increase in $\langle \hat{p}^2 \rangle$ over time, but instead $\langle \hat{p}^2 \rangle$ alternates between two values \cite{phase_noise_dkr_white_ruddell_hoogerland,BEC_Ullah,saunders_halkyard_challis_gardiner_2007,Oskay_2000}.

\section{Treating Finite-Duration Pulses\label{finite_duration}}
\subsection{Motivation for a pseudoclassical approach\label{pseudoclassical_motivation}}
In the Floquet operator for the quantum $\delta$-kicked particle the position and momentum operators are explicitly separated, making numerical determination of the system time evolution straightforward. Incorporating finite duration pulses combines $\hat{x}$ and $\hat{p}$ in the $\hat{U}_{\mathrm{Kick}}$ operator of Eq.~(\ref{floquetfiniteduration}), substantially increasing the numerical task. We are therefore motivated to introduce a simpler treatment, based on \textit{$\epsilon$-pseudoclassics}, which is intended to approximate the fully quantum treatment in an appropriate regime; similar treatments can be found in \cite{bach_burnett_d'arcy_gardiner_2005,cyrus_1997,stable_resonances_fishman,quantum_resonances_decoherence_wimberger,classical_scaling_theory_resonances_wimberger}. The evolution of a quantum particle or ensemble of quantum particles is modeled by a Monte Carlo simulation of an ensemble of pseudoclassical particles (described in section \ref{monte_carlo}), attractive both due to its computational simplicity and dynamical insight. 

\subsection{Derivation of the pseudoclassical model\label{pseudoclassical_model}}
We begin with the Floquet operator corresponding to the kicked-particle Hamiltonian, restricted to a particular $\beta$subspace [Eq.~(\ref{floquetfinitedurationB})], together with the constraint $T=\ell T_{T}/2$ (where $\ell$ is an even integer --- this corresponds to the condition for quantum resonance for the $\delta$-kicked particle). Introducing the dimensionless pulse duration $\epsilon = \hbar K^{2} t_{p}/M$, we may rewrite Eq.~(\ref{floquetfinitedurationB}) as
\begin{equation}
\begin{split}
\hat{F}(\beta) =& \exp
\left(
i\left[
\frac{\hat{k}^{2}}{2}\epsilon +\hat{k}\beta(\epsilon - 2\pi\ell)
\right]
\right)
\\& \times
\exp
\left(
-i\left[
\frac{\hat{k}^{2}}{2}\epsilon + \hat{k}\beta\epsilon
-\phi_d \cos(\hat{\theta})\right]
\right)\, .
\label{floquetfinitedurationk}
\end{split}
\end{equation}
 We now define a rescaled and shifted discrete momentum $\hat{\mathcal{J}}(\beta) = (\hat{k}+\beta)\epsilon $, leading to the commutator $[\hat{\theta},\hat{\mathcal{J}}(\beta)] = i\epsilon$. Introducing the rescaled kicking strength $\tilde{V} =\epsilon\phi_d$, we can now rewrite Eq.~(\ref{floquetfinitedurationk}) as
\begin{equation}
\begin{split}
\hat{F}(\beta) =& \exp
\left(
\frac{i}{\epsilon}\left[
\frac{\hat{\mathcal{J}}(\beta)^{2}}{2} - \hat{\mathcal{J}}(\beta)2\pi\ell \beta 
\right]
\right)
\\&\times
\exp
\left(
-\frac{i}{\epsilon}\left[
\frac{\hat{\mathcal{J}}(\beta)^{2}}{2}
-\tilde{V}\cos(\hat{\theta})\right]
\right)\, .
\end{split}
\label{floquetfinitedurationrescaled}
\end{equation}
Note that $\epsilon$ appears where we would normally expect to see $\hbar$; for small values of $\epsilon$, we therefore expect an effective classical model to give reasonable results which well approximate the quantum treatment \cite{bach_burnett_d'arcy_gardiner_2005,cyrus_1997,stable_resonances_fishman,quantum_resonances_decoherence_wimberger,classical_scaling_theory_resonances_wimberger}. 

The dynamics governed by Eq.~(\ref{floquetfinitedurationrescaled}) are equivalent to those generated by the following dimensionless Hamiltonians:
\begin{subequations}
\begin{align}
\label{H_1}
\hat{H}_{1} = & \frac{\hat{\mathcal{J}}(\beta)^{2}}{2} - \tilde{V}\cos(\hat{\theta})\, ,
\\
\label{H_2}
\hat{H}_{2} = & -\frac{\hat{\mathcal{J}}(\beta)^{2}}{2} + \hat{\mathcal{J}}(\beta)2\pi\ell\beta\, ,
\end{align}
\end{subequations}
where $\hat{H}_1$ is associated with the kick, $\hat{H}_2$ with the free evolution, and each Hamiltonian governs the time-evolution for one dimensionless time unit (rescaled time given by $t/t_{p}$). Replacing the quantum Hamiltonian $\hat{H}_1$ with its classical counterpart $H_1$, we determine Hamilton's equations of motion:
\begin{subequations}
\label{classicalmap1}
\begin{align}
\label{classicalmap1a}
\dot{\theta}& =\frac{\partial H_1}{\partial \mathcal{J}(\beta)}=\mathcal{J}(\beta)\, , \\
\dot{\mathcal{J}}(\beta)& =-\frac{\partial H_1}{\partial\theta}=-\tilde{V} \sin(\theta),
\label{classicalmap1b}
\end{align}
\end{subequations}
which we recognize as the equations of motion of a simple pendulum, the phase space orbits of which are in principle exactly solvable in terms of Jacobi elliptic functions (although they can be more convenient to solve numerically). Referring to a phase space point immediately before the $n^\mathrm{th}$ kick as $(\theta_{n},\mathcal{J}_{n}(\beta))$, we say that evolving these values under Eq.~(\ref{classicalmap1}) for 1 dimensionless time unit yields $(\theta_{n^{+}},\mathcal{J}_{n^{+}}(\beta))$. Feeding these values into the classical equations of motion generated by $H_{2}$ yields the very simple classical map
\begin{subequations}
\label{classicalmap2}
\begin{align}
\theta_{n+1} &= \theta_{n^{+}} - \mathcal{J}_{n^{+}}(\beta) + 2\pi\ell\beta, \\
\mathcal{J}_{n+1}(\beta) &=   \mathcal{J}_{n^{+}}(\beta),
\end{align}
\end{subequations}
where $(\theta_{n+1},\mathcal{J}_{n+1}(\beta))$ is the phase space point evolved to just before the $(n+1)^\mathrm{th}$ kick.

Finally, relating the dimensionless momentum $\mathcal{J}(\beta)$ back to the momentum $p$ yields:
\begin{equation}
p = \hbar K( k + \beta) 
= \frac{\hbar K}{\epsilon}\mathcal{J}(\beta).  
\end{equation}
To calculate the time evolution of expectation values using this treatment, 
we evolve an appropriate initial ensemble of classical particles and then compute their normalized statistics, as described below.

\begin{figure}[t]
{\centering\includegraphics{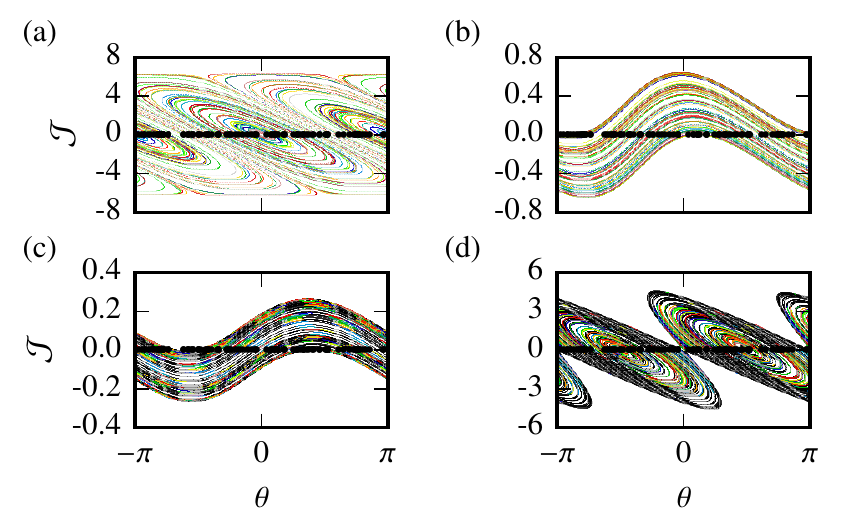}\caption{\label{fig:phasespacevaryingbeta} 
(Color online) Poincar\'{e} sections for $(\theta,\mathcal{J}(\beta))$ as evolved by Eq.~(\ref{classicalmap1}) and Eq.~(\ref{classicalmap2}),  corresponding to the $\beta=0$, 0.05, 0.2, and 0.25 subspaces for (a), (b), (c), and (d) respectively, with $\ell=2$ and $\tilde{V}=0.251$. Each black circle represents one of 100 initial phase-space points, and each color represents the evolution of a single phase-space point over 1000 kicks. The smaller black points in (c) and (d) link up the rotational or elliptic orbits, respectively, (which for $\beta = 0.2$ and $\beta=0.25$, respectively, takes substantially longer than 1000 kicks).}}
\end{figure}

\section{Monte Carlo Simulations \label{monte_carlo}}
\subsection{Quantum model}
In our finite-temperature simulations, we follow the approach of Saunders \textit{et al}.\ \cite{ saunders_halkyard_challis_gardiner_2007}, and work within the momentum basis. The initial states are momentum eigenstates, with randomly distributed values sampled from the Maxwell--Boltzmann distribution:
\begin{equation}
D_k(\beta)=\frac{1}{w \sqrt{2\pi}} \exp \left(\frac{-[k+\beta]^2}{2 w^2} \right),
\label{gaussiandist}
\end{equation}
where the temperature $\mathcal{T}_{w}=\hbar^{2} K^{2} w^{2}/M k_{\mathrm{B}}$\cite{saunders_halkyard_challis_gardiner_2007}. 

 Time-evolving an initial momentum eigenstate $|(\hbar K)^{-1}p = k + \beta\rangle$ using the Floquet operator $\hat{F}(\beta)$ of Eq.~(\ref{floquetfinitedurationk}) results in a transfer of the initial population among other momentum eigenstates, such that the time-evolved state can be written $|\psi (t) \rangle_j=\sum_k c_{kj}(t)\, |(\hbar K)^{-1}p = k + \beta\rangle$, where $|\psi (t) \rangle_j$ is the time-evolved state corresponding the the $j^\mathrm{th}$ of $N_q$ initial momentum eigenstates. The second order momentum moment is given by $\langle \hat{p}^2 \rangle(t) =N_{q}^{-1}\sum_j  \langle \hat{p}^2 \rangle_j (t)= N_{q}^{-1}\sum_j \langle \psi (t) |_j \hat{p}^2 | \psi (t) \rangle_j $. The momentum distribution can be read off from the absolute square of the coefficients $c_{kj}(t)$ for the case of a single initial momentum state, and tells us the probability of the system being in a given $k$ subspace (some given value of $k$, but any value of $\beta$). For an ensemble of $N_q$ states, the total probability $P_k(t)$ of finding an atom with a certain discrete momentum $k$ is given by the normalized sum of the absolute squares of the $c_{kj}(t)$ coefficients,  $P_k=N_{q}^{-1}\sum_j \lvert c_{kj}(t) \rvert ^2$.

\begin{figure}[t]
{\centering\includegraphics{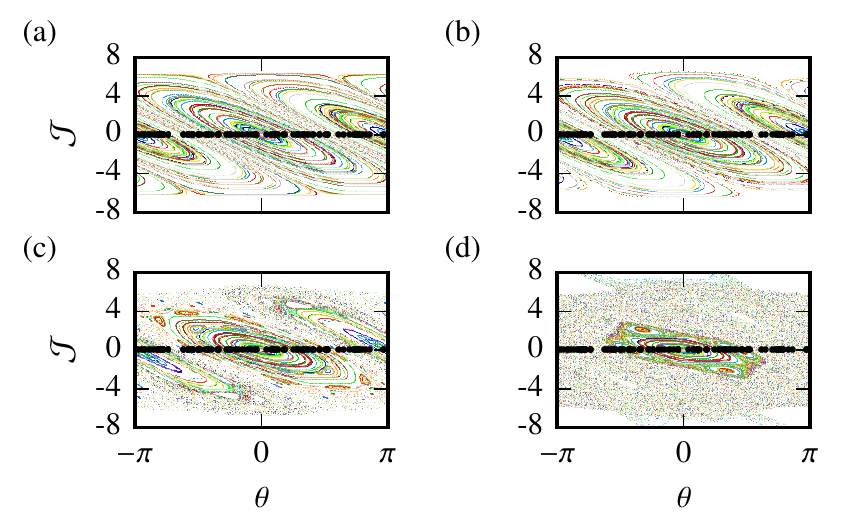}\caption{\label{fig:phasespacevaryingv}
(Color online) Poincar\'{e} sections for $(\theta,\mathcal{J}(\beta))$ as evolved by Eq.~(\ref{classicalmap1}) and Eq.~(\ref{classicalmap2}), corresponding to the $\beta=0$ subspace for driving strengths $\tilde{V}=0.251$, 2.51, 5.01, and 7.51 for (a), (b), (c), and (d) respectively, with $\ell=2$. Each black circle represents one of 100 initial phase-space points, and each color represents the evolution of a single phase-space point over 1000 kicks.}}
\end{figure}

It is desirable for our momentum distribution plots to be log-normalized so that fine features may be resolved. In practice momentum states with higher $k$-values receive a negligible amount of population compared to states near $k=0$, and so when displaying our momentum distributions we impose a cutoff value $C$, such that the condition $P_k \geqslant C$ is true for all $P_k$ and $t$ and the problem of taking the logarithm of a near-zero population is avoided.

\subsection{$\epsilon$-pseudoclassical model}
In the case of the $\epsilon$-pseudoclassical model, momentum distribution dynamics are obtained by evolving a statistical ensemble of $N_c$ classical particles according to Eq.~(\ref{classicalmap1}) and Eq.~(\ref{classicalmap2}) (note that $N_c$ need not in general be equal to $N_q$). Though the trajectory of each particle does not  in itself have a clear physical meaning, the evolution of an ensemble of sufficiently large size can be used to produce a facsimile of the quantum momentum-state-population-distribution of the gas. We place the momentum data into bins of width $\Delta p=\hbar K$, normalize the resultant population distribution and from this extract the mean squared momentum.

It is possible to produce an approximate momentum distribution also for the case of a zero temperature gas, by setting $\mathcal{J}(\beta)=0$ and choosing a random ensemble of initial $\theta$ values; the ensemble approximates a single momentum eigenstate with a given $\beta$. For the case of a finite-temperature gas, $\mathcal{J}(\beta)$ values are randomly drawn from a Maxwell--Boltzmann distribution, and $\theta$ values from a uniform distribution.

\begin{figure}[t]
{\centering
\includegraphics{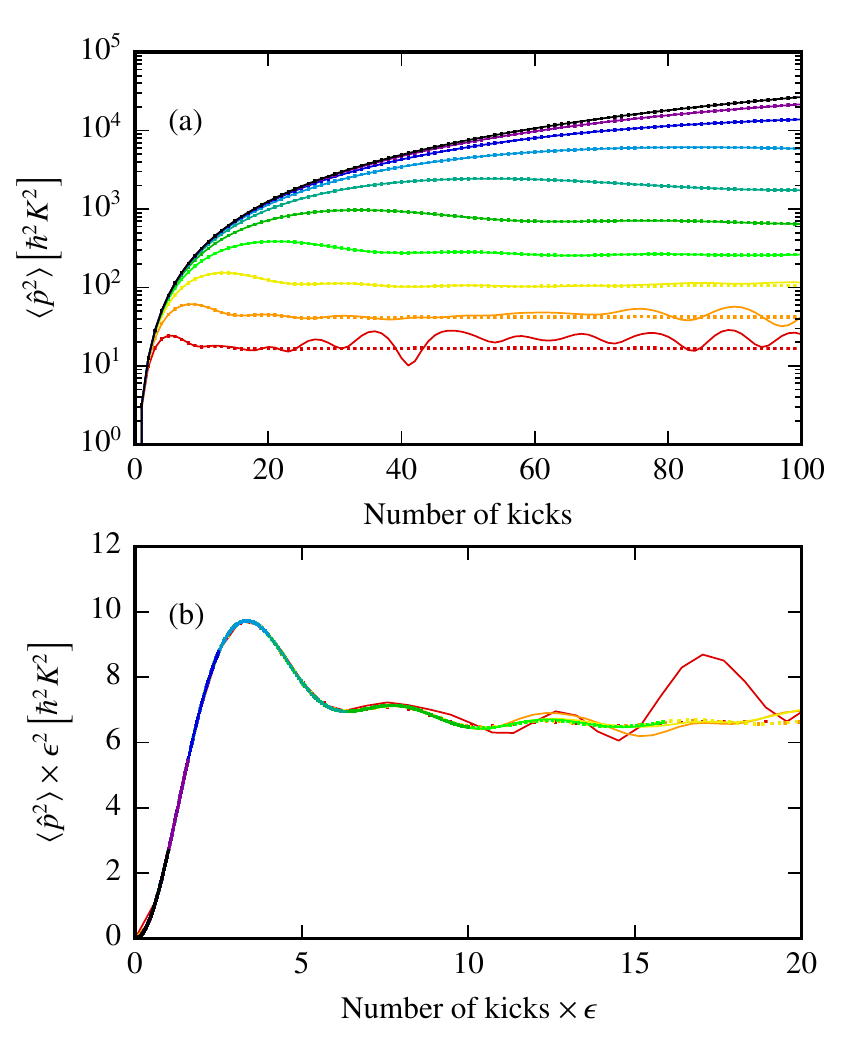}
\caption{\label{fig:epsilonvariation}(Color online) (a) Plot of $\langle \hat{p}^2 \rangle$ in units of $\hbar^2 K^2$ vs.\ number of kicks for a zero temperature gas, with $\phi_d=0.8 \pi$ and $\ell=2$. The scaled pulse duration $\epsilon$ takes the values $10^{-2 +2 j/11}$, where $j=\left\{0,1,2,..,10\right\}$. The curves represent the results of the quantum dynamics [Eq.~(\ref{floquetfiniteduration})], and the points those of the $\epsilon$-pseudoclassical model [Eq.~(\ref{classicalmap1}) and Eq.~(\ref{classicalmap2})], with lower values of $\epsilon$ giving rise to higher peak values of $\langle \hat{p}^2 \rangle$. Hence, the black curve corresponds to $\epsilon=0.01$ ($j=0$) and the red curve to $\epsilon=0.658$ ($j=10$). (b) Rescaling of (a) by $\epsilon^2$ in the $\langle \hat{p}^2 \rangle$ axis and $\epsilon$ in the kick-number axis such that a universal curve is revealed, where all data overlap over a suitably short timescale.}}
\end{figure}

\begin{figure*}[htp]
{\centering
\subfloat{\includegraphics{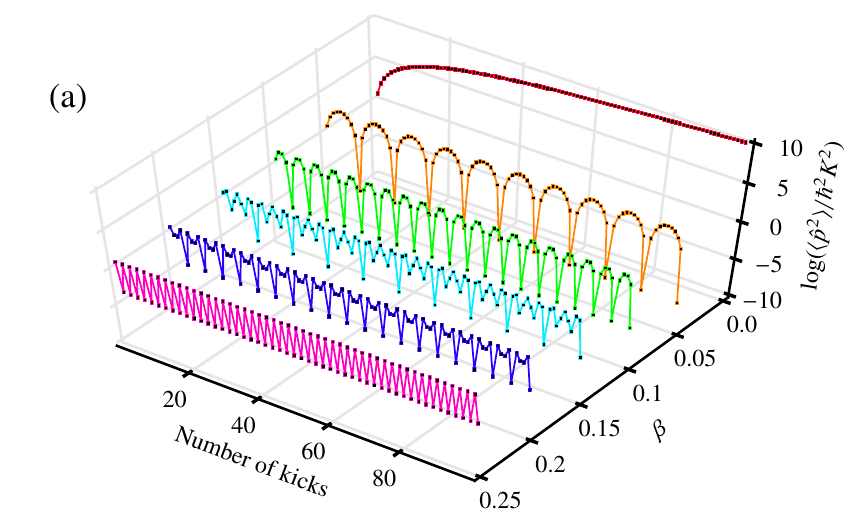}}\quad
\subfloat{\includegraphics{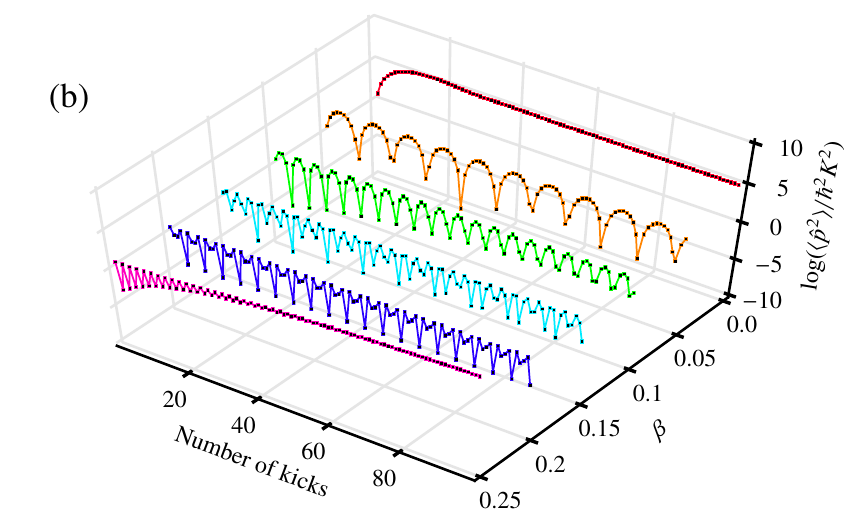}}
\caption{\label{betadependencewaterfall}(Color online) Plots of the time evolution of the log of $\langle \hat{p}^2 \rangle$ in units of $\hbar^2 K^2$ vs.\ number of kicks, for different values of the quasimomentum $\beta =\{0,0.05,0.1,0.15,0.2,0.25\}$, for an otherwise zero temperature gas [initial momentum eigenstate with $\mathcal{J}(\beta)=0$]. The smooth curves represent results of the quantum evolution [Eq.~(\ref{floquetfiniteduration})], and the points those of the effective classical model [Eq.~(\ref{classicalmap1}) and Eq.~(\ref{classicalmap2})]. For figure (a) $\epsilon=0.001$, and for figure (b) $\epsilon=0.2$. Other parameters are $\phi_d=0.8 \pi$  and $\ell=2$.}}
\end{figure*}

\section{Results\label{results}}
\subsection{Dynamics of the pseudoclassical map \label{pseudoclassical_dynamics}}
To gain insight into the system dynamics, it is useful to construct $(\theta,\mathcal{J})$ Poincar\'{e} sections, which in this case are stroboscopic maps defined by Eq.~(\ref{classicalmap1}) and Eq.~(\ref{classicalmap2}), evolved for some number of kicks $N$. We remark that we have opted to solve the equations of motion generated by $H_1$ numerically rather than using the exact Jacobi elliptic functions for ease of implementation; this still requires vastly less computational power to solve the time evolution of the system than the Fourier methods generally used in a fully quantum treatment. Inspection of Eq.~(\ref{classicalmap1}) and Eq.~(\ref{classicalmap2}) reveals that there are exactly two free parameters: the driving strength $\tilde{V}$, and the quasimomentum $\beta$. We therefore construct a selection of Poincar\'{e} sections varying these, choosing $\tilde{V}=0.251$ when we vary $\beta$ (Fig.~\ref{fig:phasespacevaryingbeta} --- this value is motivated by typical experimental values \cite{BEC_Ullah,Rebuzzini_2005_quantum_resonance,Schlunk_2003_accelerator_modes_stability,Schlunk_2003_accelerator_modes,Ma_2004_accelerator_modes,Buchleitner_2006_accelerator_modes,Oberthaler_1999_accelerator_modes,godun_2000_accelerator_modes,darcy_2001,darcy_2001_momentum_diffusion,Darcy_2003_accelerator_modes}), and $\beta=0$ when we vary $\tilde{V}$ (Fig.~\ref{fig:phasespacevaryingv}).

The Poincar\'{e} section of Fig.~\ref{fig:phasespacevaryingbeta}(a) [repeated  in Fig.~\ref{fig:phasespacevaryingv}(a) for ease of comparison between different $\beta$ subspaces and values of $\tilde{V}$] corresponds to that of an exact quantum resonance in the $\delta$-kicked particle case (for which the dynamical behaviour varies from resonant to antiresonant, depending on the value of $\beta$  \cite{saunders_halkyard_challis_gardiner_2007,saunders_halkyard_gardiner_challis_2009,power_law_behaviour_Halkyard}). There are two stable fixed points visible at (0,0) and $(-\pi,0) \equiv (\pi,0)$, each 
surrounded by concentric orbits characteristic of regular (non-chaotic) motion. Fig.~\ref{fig:phasespacevaryingbeta} (d) corresponds to the $\beta=0.25$ subspace, which we expect to behave as an  antiresonance in the $\delta$-kicked limit. Clearly the system dynamics vary dramatically between different $\beta$ subspaces, and we must therefore consider them all when modeling a thermal gas.

In Fig.~\ref{fig:phasespacevaryingv} we see that, as we increase the driving strength $\tilde{V}$ from $\tilde{V}=0.251$, a region of pseudorandom trajectories opens up in the outer parts of each system of elliptic orbits, until the Poincar\'{e} section becomes predominantly chaotic for $\tilde{V}=7.51$. We remark that such high values of $\tilde{V}$, combined with small values of $\epsilon$, correspond to very high laser intensities, making it unclear what the transition to chaos in the $\epsilon$-pseudoclassical model really represents in an atom-optical context.

\subsection{Zero-temperature gas\label{zero_temp_comparison}}
We now compute the evolution of $\langle \hat{p}^2 \rangle$ over time for a range of values of $\epsilon$ and constant $\phi_{d}$ (meaning that $\tilde{V}\equiv\epsilon\phi_{d}$ scales linearly with $\epsilon$), using both the pseudoclassical and fully quantum calculations, at zero temperature. This is actually computationally straightforward in the quantum case, as one need only evolve a single initial (zero momentum) eigenstate. 

We display our results in Fig.~\ref{fig:epsilonvariation}(a). Two behaviors are clearly visible: 
\begin{enumerate}
\item As $\epsilon$ increases, the approximate pseudoclassical simulations deviate from the quantum dynamics after a smaller number of kicks. As this model relies on an expansion about $\epsilon$ as a smallness parameter, this deviation can be thought of as a cumulative error in the pseudoclassical dynamics that increases in magnitude each time the classical maps are applied. Results like those of Fig.~\ref{fig:epsilonvariation}(a) allow us to characterize time scales  over which we can expect agreement between the pseudoclassical and quantum treatments for a given value of $\epsilon$.

\begin{figure*}[!t]
{\centering
\includegraphics{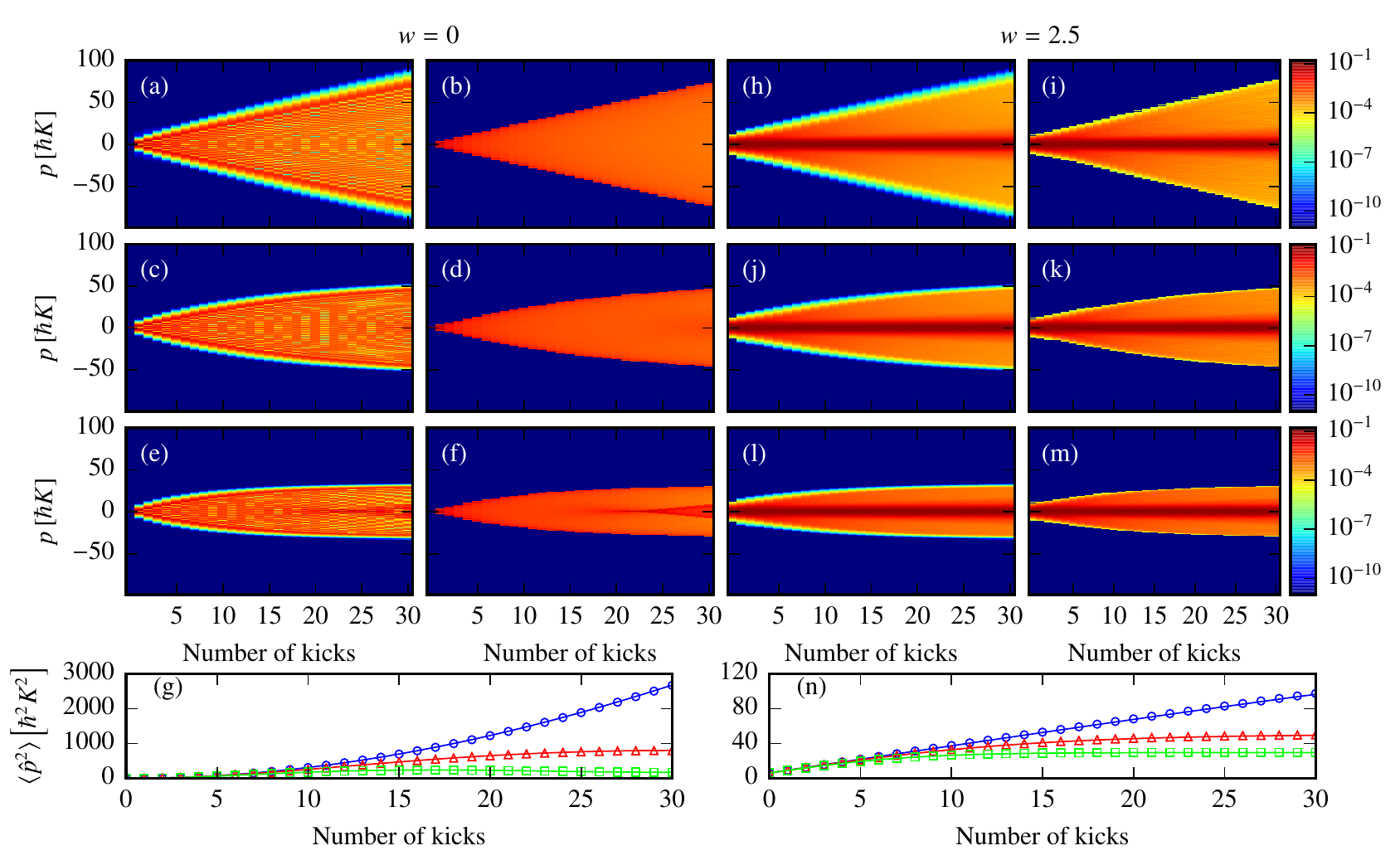}
\caption{\label{momentumdistributionsnotr}
(Color online) Comparison between the dynamics of the momentum distributions computed by the fully quantum model, [Eq.~(\ref{floquetfiniteduration})], and the pseudoclassical model [Eq.~(\ref{classicalmap1}) and Eq.~(\ref{classicalmap2})] for zero ($w=0$) and finite temperature gases ($w=2.5$), with $\phi_d=0.8 \pi$ and $\ell$=2, for differing values of the scaled pulse duration $\epsilon$. The first and second columns show momentum distributions for a zero temperature gas ($w=0$) as computed by the quantum [(a), (c), (e)] and pseudoclassical models [(b), (d), (f)] respectively. Columns 3 and 4 give the momentum distributions computed by the quantum [(h), (j), (l)] and effective classical models  [(i), (k), (m)] respectively, for $w=2.5$. In each row, the distribution dynamics are computed for a different value of $\epsilon$: row 1 [(a), (b), (h), (i)] has $\epsilon=0.02$, row 2 [(c), (d), (j), (k)] has $\epsilon=0.11$,  and row 3 [(e), (f), (l), (m)] has $\epsilon=0.2$. To accommodate the logarithmic color scale, we have chosen a cutoff value of $C=10^{-11}$. The corresponding time-evolution of $\langle \hat{p}^2 \rangle$ [in units of $\hbar^{2} K^{2}$] is given in (g), for $w=0$ and (n) for $w=2.5$; solid lines represent results of quantum calculations, and symbols those of the effective classical model (squares correspond to $\epsilon=0.2$, triangles to $\epsilon=0.11$, and circles to $\epsilon=0.02$). Monte Carlo calculations were carried out with $N_{c}=10^5$ particles, or $N_{q}=10^5$ state vectors, as appropriate.}}
\end{figure*}

\item The peak value of $\langle \hat{p}^2 \rangle$ is higher for smaller values of $\epsilon$. Recalling that $\epsilon$ is simply a rescaled pulse duration, as it approaches zero the system behaves increasingly as if it were receiving $\delta$-kicks, for which $\langle \hat{p}^2 \rangle$ would increase indefinitely over time.  It is again clear that the smaller the value of $\epsilon$, the longer the timescale over which the system behaves as if it were $\delta$-kicked. At an $\epsilon$-dependent point in time, $\langle \hat{p}^2 \rangle$ deviates from the quadratic growth associated with perfect quantum resonance, corresponding to violation of the Raman--Nath regime. We can see that $\langle \hat{p}^2 \rangle$ must eventually decrease by inspection of the phase-space diagram in Fig.~\ref{fig:phasespacevaryingbeta}(a), as the spread of trajectories is forced to eventually decrease simply because they manifest as bounded quasiperiodic orbits.
\end{enumerate}

Rescaling the axes in Fig.~\ref{fig:epsilonvariation}(a) according to the value of $\epsilon$ reveals a universal curve, which exists independent of this value, as displayed in Fig.~\ref{fig:epsilonvariation}(b). 
This universality appears to be essentially exact in the pseudoclassical model, but ceases to apply for the quantum calculations once they deviate significantly from the pseudoclassical predictions.
The observed oscillating decay encapsulates the dynamics visible in Fig.~\ref{fig:phasespacevaryingbeta}(a), and appears indicative of the dephasing of an ensemble of anharmonic oscillators.

Figure~\ref{betadependencewaterfall} shows comparisons of $\langle \hat{p}^2 \rangle$ evolution as computed by the quantum and $\epsilon$-pseudoclassical models for initial conditions corresponding to a single momentum eigenstate with $k=0$ and different values of $\beta$. The pseudoclassical and quantum models agree well over the entire range of $\beta$ subspaces. Hence, for any reasonable initial momentum distribution, we can expect the pseudoclassical model to reproduce the correct quantum dynamics provided that $\epsilon$ is small enough on the timescale to be considered. We have chosen $\epsilon=0.001$ for Fig.~\ref{betadependencewaterfall}(a), where the dynamics are essentially coincident with those induced by perfect $\delta$-kicks for the chosen parameters and kick numbers. In Fig.~\ref{betadependencewaterfall}(b) we have $\epsilon=0.2$; comparing with Fig.~\ref{betadependencewaterfall}(a) it is clear that time evolution of $\langle \hat{p}^2 \rangle$ is significantly affected by the finite duration of the kicking pulses. Note, however, that although $\epsilon=0.2$ would seem to be borderline in terms of being a ``small parameter,'' the agreement between the $\epsilon$-pseudoclassical model and the full quantum dynamics still appears to be excellent.

\begin{figure*}[!ht]
{\centering
\includegraphics{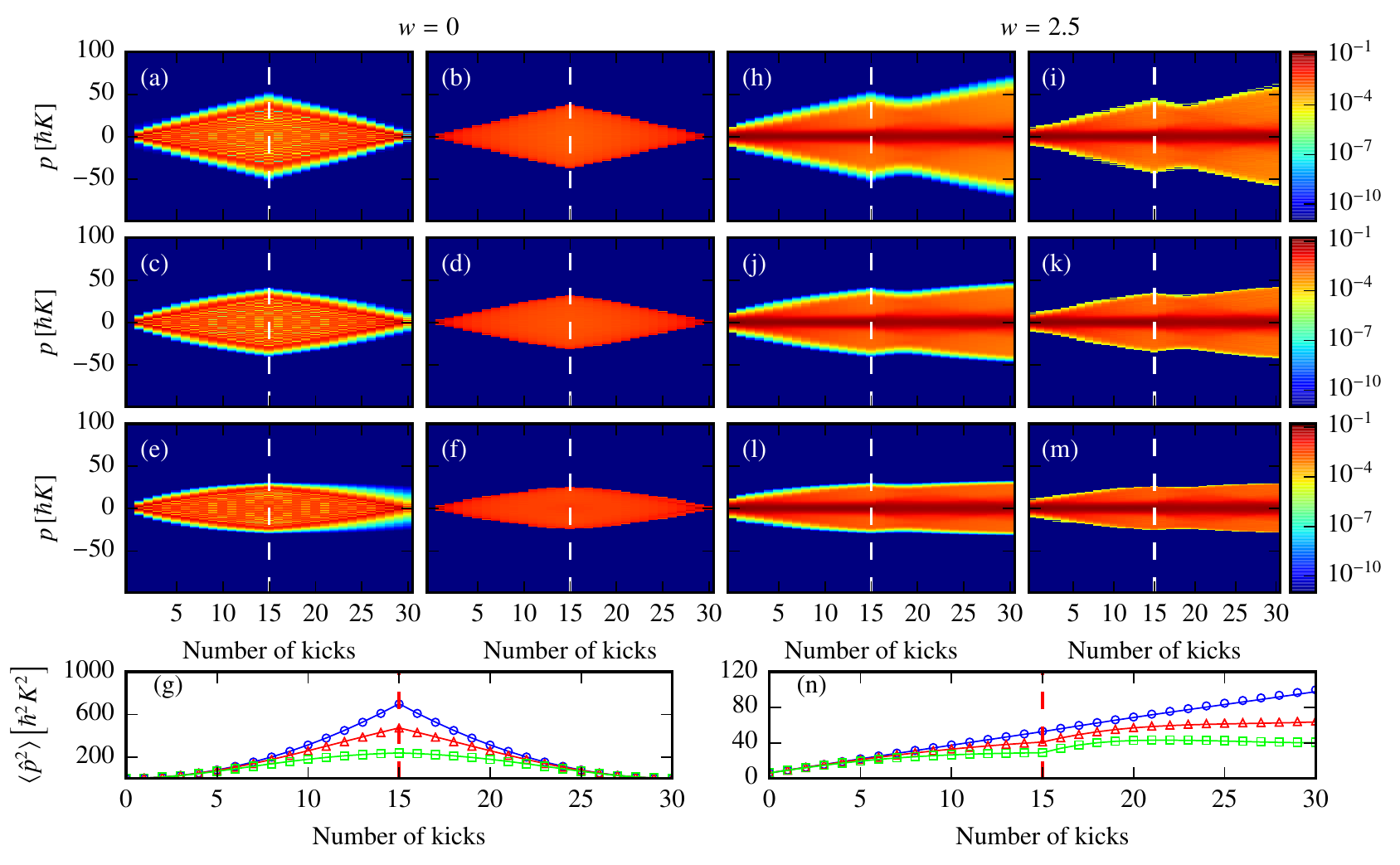}
\caption{\label{momentumdistributions}
(Color online) Comparison between the dynamics of the momentum distributions computed by the fully quantum model, [Eq.~(\ref{floquetfiniteduration})], and the pseudoclassical model [Eq.~(\ref{classicalmap1}) and Eq.~(\ref{classicalmap2})] for zero ($w=0$) and finite temperature gases ($w=2.5$), with $\phi_d=0.8 \pi$ and $\ell$=2, for differing values of the scaled pulse duration $\epsilon$. In each case a time-reversal event (phase-shifitng the standing wave by $\pi$) occurs at the $15^{\text{th}}$ of 30 kicks (marked by the dashed lines). The first and second columns show momentum distributions for a zero temperature gas ($w=0$) as computed by the quantum [(a), (c), (e)] and pseudoclassical models [(b), (d), (f)] respectively. Columns 3 and 4 give the momentum distributions computed by the quantum [(h), (j), (l)] and effective classical models  [(i), (k), (m)] respectively, for $w=2.5$. In each row, the distribution dynamics are computed for a different value of $\epsilon$: row 1 [(a), (b), (h), (i)] has $\epsilon=0.02$, row 2 [(c), (d), (j), (k)] has $\epsilon=0.11$,  and row 3 [(e), (f), (l), (m)] has $\epsilon=0.2$. To accommodate the logarithmic color scale, we have chosen a cutoff value of $C=10^{-11}$. The corresponding time-evolution of $\langle \hat{p}^2 \rangle$ [in units of $\hbar^{2} K^{2}$] is given in (g), for $w=0$ and (n) for $w=2.5$; solid lines represent results of quantum calculations, and symbols those of the effective classical model (squares correspond to $\epsilon=0.2$, triangles to $\epsilon=0.11$, and circles to $\epsilon=0.02$). Monte Carlo calculations were carried out with $N_{c}=10^5$ particles, or $N_{q}=10^5$ state vectors, as appropriate.}}
\end{figure*}

As $\beta$ increases from $0$ the evolution of $\langle \hat{p}^2 \rangle$ over time progresses from resonant to antiresonant behavior. This progression is twofold periodic in the space of quasimomenta: Eq.~(\ref{classicalmap2}) shows that for $\ell = 2$ the same pseudoclassical dynamics are observed for $\beta+1/2$ as for $\beta$ (this symmetry can also be deduced for expectation values derived from the fully quantal Floquet operator [Eq.~(\ref{floquetfinitedurationrescaled})] acting on momentum eigenstates \cite{saunders_halkyard_challis_gardiner_2007}). Furthermore, the Hamiltonian is an even function of both $\hat{p}$ and $\hat{x}$, meaning that the same $\langle\hat{p}^{2}\rangle$ dynamics are observed for $-\beta$ as for $\beta$. Hence, the data plotted in Fig.~\ref{betadependencewaterfall} effectively span the full range of $\beta$ dependencies when the initial value of $\mathcal{J}(\beta)$ (or $k$) is equal to $0$.

\subsection{Finite-temperature Monte Carlo \label{monte_carlo_comparison}}
We now perform comparative quantum and pseudoclassical Monte Carlo simulations for experimentally achievable timescales. The initial finite temperature ensembles are chosen by random sampling from a Maxwell--Boltzmann distribution (combined with a uniform distribution for $\theta$ in the case of the pseudoclassical dynamics), as described in section~\ref{monte_carlo}. In Figs.~\ref{momentumdistributionsnotr}(a--f) and Figs.~\ref{momentumdistributionsnotr}(h--m), we compare momentum distributions, computed for three values of $\epsilon$, using both the pseudoclassical and quantum treatments, over a small number of kicks, at zero temperature ($w=0$) and for Cesium atoms at $\mathcal{T}_{w}\simeq 5\,\mu$K ($w=2.5$). In Fig.~\ref{momentumdistributionsnotr}(g)  and Fig.~\ref{momentumdistributionsnotr}(n) we show the associated values of $\langle \hat{p}^2 \rangle $ computed for each case to check that our comparison takes place within the regime of validity of the $\epsilon$-pseudoclassical model. For the zero-temperature ($w=0$) case, the population splitting in momentum space characteristic of a quantum resonance can be seen in both models over the full $30$ kicks for $\epsilon=0.02$. For larger values of $\epsilon$ we observe a slowing in the momentum spreading, followed by a clear plateau in the case of $\epsilon = 0.2$, which is also visible in the corresponding plot of $\langle \hat{p}^2 \rangle $.  

For each value of $\epsilon$ the overall shape of the momentum distribution computed by the $\epsilon$-pseudoclassical model matches that of the fully quantum calculation well. A degree of internal structure is present in the zero-temperature ($w=0$) quantum distributions that is not present in their $\epsilon$-pseudoclassical counterparts. Similarly, in both the $w=0$ and $w=2.5$ quantum distributions, there is further structure visible, where the most extreme populated states in momentum space meet the near zero-population background, that is not present in the pseudoclassical calculation. We can clearly see from Fig.~\ref{momentumdistributionsnotr}(g) and Fig.~\ref{momentumdistributionsnotr}(n) that the evolution of $\langle \hat{p}^2 \rangle$ is nonetheless reproduced perfectly over a short time-scale. For the $w=2.5$ case, we see a clearly defined feature centered around $p=0$ representing a large concentration of population. This is typical of finite-temperature quantum-resonant dynamics in atom-optical systems \cite{saunders_halkyard_challis_gardiner_2007,Moore_1995_atom_optics_realization,Oskay_2000}, and can be understood from Fig.~\ref{betadependencewaterfall}; essentially a broad initial momentum distribution means that both resonant ($\beta=0$) and bounded antiresonant ($\beta=0.25$) dynamics take place simultaneously, as well as the whole range of intermediate behavior, leading to an overall averaging of the spreading in momentum space.

With atom interferometry in mind, we have repeated these simulations with the addition of a time-reversal event occurring at $n_R=15$ kicks (as described in section \ref{quantum_resonance}), displaying our results in Fig.~\ref{momentumdistributions}. In Fig.~\ref{momentumdistributions}(a) and Fig.~\ref{momentumdistributions}(b) ($\epsilon=0.02$ and $w=0$) we clearly have a near-perfect time-reversal process, with the majority of the population returning to the zero-momentum state when $n=2n_R$. Increasing $\epsilon$ to 0.11, we can see from Fig.~\ref{momentumdistributions}(c) and Fig.~\ref{momentumdistributions}(d) that the asymmetry about $n=2n_R$ has increased very slightly, and for $\epsilon=0.2$ we can see from Fig.~\ref{momentumdistributions}(e) and Fig.~\ref{momentumdistributions}(f) that the asymmetry has become even larger (similar effects were observed in \cite{daszuta_andersen_2012}). For $w=2.5$, however [Figs.~\ref{momentumdistributions}(h--n)], each distribution begins to refocus but subsequently increases in breadth (this is the same behaviour as expected for a $\delta$-kicked atomic gas). Note that as the value of $\epsilon$ increases the final distributions become narrower, which is an effect of using finite-duration pulses.

In each case the $\epsilon$-pseudoclassical predictions give good agreement with the shapes of the momentum distributions yielded by a fully quantum treatment, with the missing edge detail around each quantum distribution only manifest at around the $P_k=10^{-7}$ level. Crucially, it is clear that the lack of internal structure in the $\epsilon$-pseudoclassical distributions is not a problem for calculating $\langle \hat{p}^2 \rangle$ under time reversal or at finite temperature. An interferometric measurement would look at deviations from a perfect time reversal, potentially motivating a study of the fidelity of a time-reversed kicked gas with finite-duration pulses, for example using a similar approach to that derived for the $\delta$-kicked rotor in \cite{pseudoclassical_fidelity}.

\begin{figure}[!t]
{\centering
\includegraphics{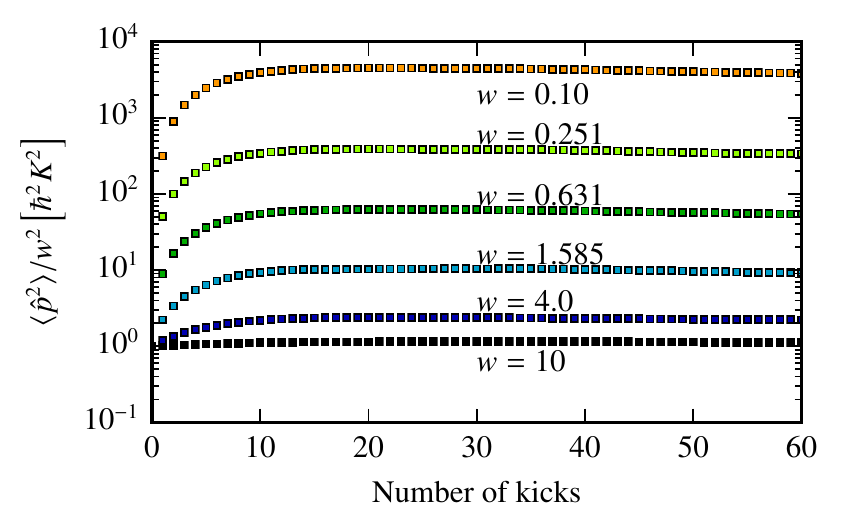}
\caption{\label{fig:tempdependence}
(Color online) Plots of the time evolution of $\langle \hat{p}^2 \rangle/w^2$ in units of $\hbar^2 K^2$ vs.\ number of kicks, with $\epsilon=0.2$, $\phi_d=0.8 \pi$ and $\ell=2$. Each set of points corresponds to an individual value of $w = 10^{-1 +2 j/5}$, where $j=\left\{0,1,2,...,5\right\}$, as computed by the pseudoclassical model [Eq.~(\ref{classicalmap1})  and Eq.~(\ref{classicalmap2})].}}
\end{figure}

Having carried out a detailed comparison of the quantum and $\epsilon$-pseudoclassical models over relatively short time scales and at finite temperature, we can reasonably assume that whatever value we select for $w$, the pseudoclassical model will produce accurate results, provided an appropriate value of $\epsilon$ is chosen. To better understand the variation of $\langle \hat{p}^2 \rangle$ with temperature over longer time scales, we have carried out simulations for six values of $w$, using only the $\epsilon$-pseudoclassical model (results displayed in Fig.~\ref{fig:tempdependence}). We choose $\epsilon=0.2$ for each simulation, as this is a relatively large value where we have already shown excellent agreement in $\langle \hat{p}^2 \rangle$ with the fully quantum treatment over a range of 100 kicks (see Fig.~\ref{betadependencewaterfall}). Plotting $\langle \hat{p}^{2}\rangle/w^{2}$ versus the number of kicks $n$, the $n=0$ value for each curve is the same, but from $n=1$ they separate markedly --- the lower the value of $w$, the greater the relative increase, due to the increased dominance of quantum-resonant behavior centered at $\beta=0$. The computational simplicity of the pseudoclassical model means that such a plot can be produced in a few minutes on a standard desktop computer, which is potentially invaluable when planning a hypothetical atom-interferometry experiment.

\section{Conclusions\label{conclusions}}
We have derived an $\epsilon$-pseudoclassical model for quantum resonances in a finite-temperature dilute atomic gas driven by finite-duration off-resonant laser pulses, and compared to its fully quantum counterpart. Dynamics of the $\epsilon$-pseudoclassical model have been investigated and certain phase space features associated with quantum resonant behavior have been identified. Further, it has been shown how increasing the parameter $\epsilon$ shortens the time-scale over which the quantum and $\epsilon$-pseudoclassical calculations agree at zero temperature, as well as the amount of time before a quantum resonance begins to plateau due to violation of the Raman--Nath regime. The accuracy of the $\epsilon$-pseudoclassical model was shown to be unaffected by the initial state's quasimomentum, and is therefore suitable for treating a finite-temperature gas. Monte Carlo simulations were explicitly performed to this end and compared both the expectation value $\langle \hat{p}^2 \rangle$ and momentum distributions as computed by each model, and it was found that the $\epsilon$-pseudoclassical model reproduces the former essentially exactly, even at finite temperature, and the general shape of the latter up to small details. We have also shown explicitly that the $\epsilon$-pseudoclassical model correctly treats the time-reversal mechanism necessary for light-pulse atom-interferometry. Finally, $\epsilon$-pseudoclassical Monte Carlo simulations were performed to determine the behavior of $\langle \hat{p}^2 \rangle$ at different values of $w$ for a large number of kicks. We expect this approach to be useful in quantifying the suitability of particular experimental parameter regimes for light-pulse atom interferometry.

The data presented in this paper are available. See Ref. \cite{data}.
\acknowledgments
BTB, IGH, and SAG thank the Leverhulme Trust for support. MFA additionally thanks NZ-MBIE contract No.\ UOOX1402. BTB also thanks John L. Helm, Hannah Goodsell, Thomas P. Billam and Matthew P. A. Jones for helpful discussions.

\appendix
\section{Numerical methods \label{numerical_methods}}
For every simulation using the $\epsilon$-pseudoclassical model, Eq.~(\ref{classicalmap1}) was integrated numerically using Adams' method, as implemented in the Python module \textit{scipy.integrate.odeint}, which is based on the routine \textit{lsoda}, from the FORTRAN library \textit{odepack}. The integration time-step in the interval between kicks was adaptively variable. Convergence was checked automatically, and is also clearly indicated by the smooth nature of the phase space trajectories observed in the non-chaotic regime. The map given by Eq.~(\ref{classicalmap2}) was applied using simple matrix multiplication.
For the quantum calculations, we employed a second order split-step Fourier method, which we implemented in Python using the \textit{numpy.fft.fft} and \textit{numpy.fft.fftshift} routines. A total of 1000 split-steps were used for each kicking pulse.

\bibliography{AnEpsilonPseudoclassicalModelForQuantumResonancesinaPeriodicallyLaser-DrivenDiluteAtomicGas}

\end{document}